\documentclass[a4paper,fleqn,usenatbib]{mnras}

\usepackage{mathptmx}

\usepackage[T1]{fontenc}
\usepackage{ae,aecompl}

\usepackage{graphicx}	
\usepackage{amsmath}	
\usepackage{amssymb}	

\newcommand{\kms}{\,km\,s$^{-1}$} 

\newcommand\revchanges[1]{#1}  

\usepackage{color}  
\definecolor{myred}{rgb}{0.7,0.0,0.2}

\title[AG~Pegasi in outburst]{AG~Pegasi -- now a classical symbiotic star 
in outburst?\thanks{Based on observations obtained in Rozhen Observatory, European 
Southern Observatory program 074.D-0114 and Asiago Observatory}}

\author[T.V.~Tomov et al.]{
T.V.~Tomov,$^{1}$\thanks{E-mail: tomov@umk.pl (TT)}
K.A.~Stoyanov,$^{2}$
R.K.~Zamanov$^{2}$
\\
$^{1}$Centre for Astronomy, Faculty of Physics, Astronomy and Informatics, Nicolaus 
Copernicus University, Grudziadzka 5, 87-100 Torun, Poland\\
$^{2}$Institute of Astronomy and National Astronomical Observatory, Bulgarian Academy of 
Sciences, Tsarigradsko Shose 72, 1784 Sofia, Bulgaria\\}

\date{Accepted XXX. Received YYY; in original form ZZZ}

\pubyear{2016}

\begin{document}
\label{firstpage}
\pagerange{\pageref{firstpage}--\pageref{lastpage}}
\maketitle

\begin{abstract}

Optical spectroscopy study of the recent AG~Peg outburst observed 
during the second half of 2015 is presented. Considerable variations of the 
intensity and the shape of the spectral features as well as the changes of the hot 
component parameters, caused by the outburst, are discussed and certain similarities 
between the outburst of AG~Peg and the outburst of a classical symbiotic stars are 
shown. It seems that after the end of the symbiotic nova phase, AG~Peg became a member of 
the classical symbiotic stars group.

\end{abstract}




\section{Introduction}
\label{intro}

AG~Peg is the oldest symbiotic nova which attracted astronomers attention 
with the first spectral observations performed by \citet{1894AstAp..13..501F}. 
\citet{1921AN....213...93L} showed that in the mid of 19th century, when a brightness 
increase started, the star's visual magnitude was $\sim9$. The maximum at about 
$6^\mathrm{m}$ was reached around the year 1870. The subsequent, gradual decrease of the 
star brightness, reaching the level observed before the outburst, lasted for about 130 
years.

During this time AG~Peg demonstrated a unique evolution, passing through 
various phases: from a Be star around the maximum, through a P~Cyg-type object, to a 
symbiotic star revealing the spectra of a hot Wolf-Rayet component and cool M3 giant. All 
these changes were studied in great detail by \citet{1916POMic...2...71M, 
1929ApJ....69..330M, 1929CMWCI.381....1M,1942ApJ....95..386M, 1951ApJ...114..338M, 
1951ApJ...113..605M, 1959ApJ...129...44M}, \citet{1940ApJ....91..546S, 
1942ApJ....95..152S}, \citet{1967SvA....10..783B,1967SvA....11....8B}, 
\citet{1972PASP...84..240H}, \citet{1973ApJ...184..687C}, \citet{1975ApJ...201..404H}, 
\citet{1979ApJ...229..994G}. 

Merrill has found an 800 days long period of variation of intensities of 
emission lines and radial velocities. UBV photometric observations 
by \citet{1968SvA....12..110B, 1970Ap......6...22B, 1985IBVS.2697....1B} had confirmed 
this periodicity. Observations of AG~Peg in X-ray, UV, IR and radio ranges allowed a more 
precise evaluation of basic parameters of the system components to be made. Emergence and 
gradual attenuation of high-speed, massive and dense wind from the system's hot component 
as well as signs of its interaction with the wind from the red giant were found. Also the 
existence of a complex structure of a circumbinary nebula was revealed \citep[][and 
references therein]{1991ApJ...366..549K,2007ApJ...662.1231K, 1993AJ....106.1573K, 
2001AJ....122..349K, 1994A&A...284..145V,1994A&A...282..586M, 1995A&A...293L..13N, 
1995A&A...297L..87M}. \citet{2000AJ....119.1375F} used precise infrared radial velocities 
to determine the single-lined spectroscopic orbit for AG~Peg. Recent observations show 
that the gradual brightness decline ceased and the present brightness shows only 
orbitally related wave-like variations around the mean brightness 
\citep{2012AN....333..242S}.

As pointed out by \citet{1993AJ....106.1573K}, most authors suggest a nearly constant 
rate of the slow AG~Peg brightness decline. However, in the literature there 
are several reports of abrupt changes of the star brightness, such as a minimum of 
1907-1908 about $1^\mathrm{m}$ deep and a ``flare'' in 1946 with an amplitude about 
$1^\mathrm{m}$ \citep[see][]{1963SvA.....6..483A, 1968SvA....12..110B}. According to 
\citet{1993AJ....106.1573K}, there is nothing unusual in these brightness variations and 
they note that the ephemeris by \citet{1985PASP...97..653F} predicts both --
the minimum and the maximum. Another episode of AG~Peg UBVRI minimum brightness during 
1985, together with remarkable emission lines variations, is reported and shortly 
discussed by \citet{1987Ap&SS.131..775B}.

In this paper, we report on the AG~Peg outburst with two maxima, 
which occurred in the second half of 2015. These brightness maxima, in comparison with 
the reported before, are clearly different and in many respects are very similar to 
the multi-maxima outbursts, typical for the classical symbiotic stars like AG~Dra and 
Z~And.

\section{Observations and data reduction}
\label{observ}

\subsection{Photometry}
\label{photom}

\begin{figure}
	\includegraphics[width=\columnwidth]{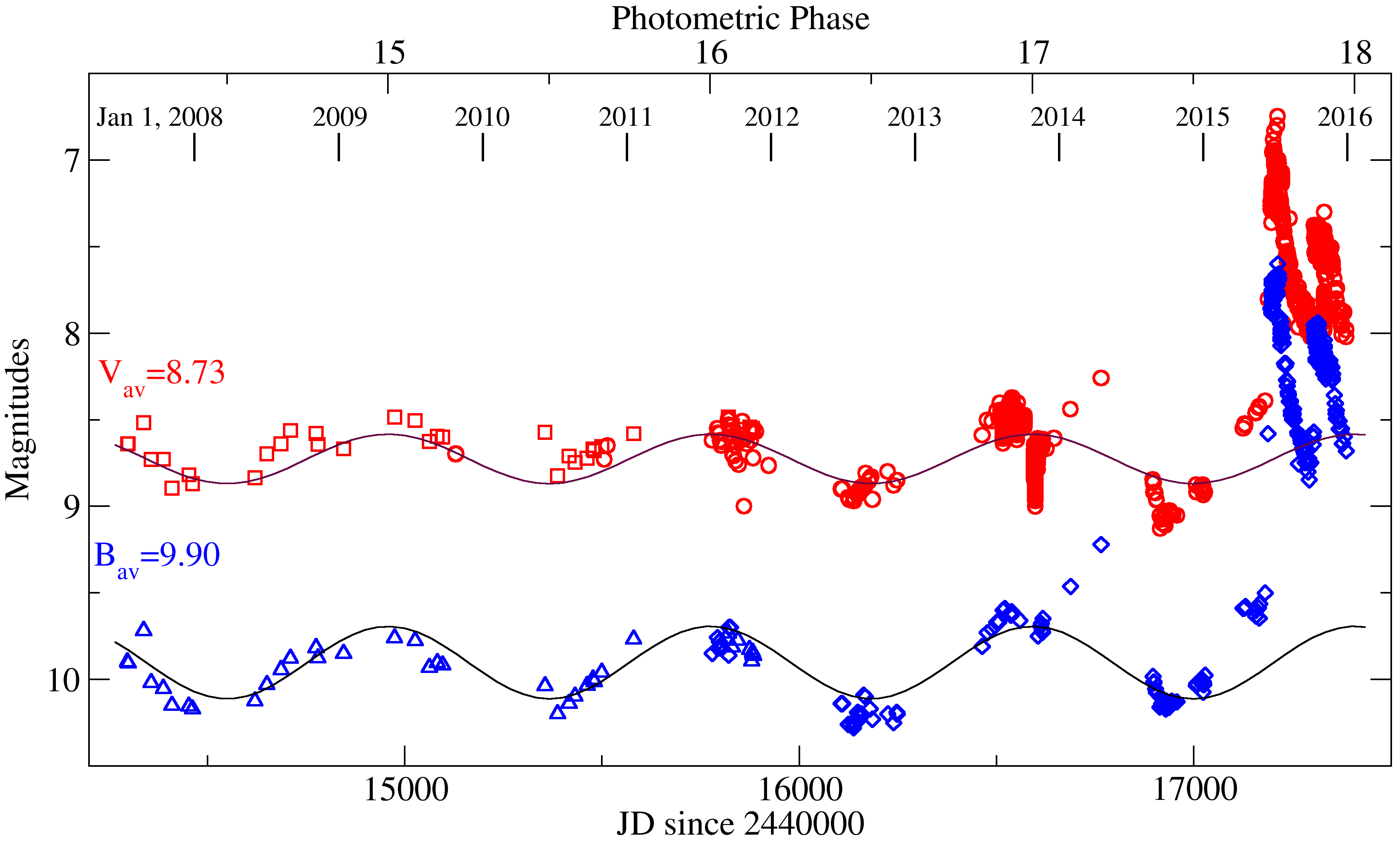}
    \caption{B and V light curves of AG~Peg from mid 2007 to the end of 2015. The data 
presented with \textsl{squares} and \textsl{triangles} is from \citet{2012AN....333..242S} 
while presented with \textsl{circles} and \textsl{diamonds} is from AAVSO \citep{kafka}. 
The solid lines represent the fit to the data using \citet{1985PASP...97..653F} 
ephemeris.}
    \label{fig:lc}
\end{figure}

The AG~Peg light curves in V and B bands, from the period of the mid of 2007 
to the end of 2015, based on data published by \citet{2012AN....333..242S} and the AAVSO 
database \citep{kafka}, are shown in Fig.~\ref{fig:lc}. Following 
\citet{1993AJ....106.1573K,2001AJ....122..349K} we used the \citet{1985PASP...97..653F} 
ephemeris for maximum visual light
\begin{equation}
\mathrm{Max}(V) = \mathrm{JD}\,2442710.1+816.5\mathrm{E}
\end{equation} to calculate the photometric phase and to fit the brightness 
variations. Apparently, the AG~Peg brightness changes are consistent with this 
ephemeris around mean values of 9\fm90 and 8\fm73 for B and V respectively. 
\citet{2013ATel.5258....1M} announced increased star activity around photometric phases 
16.8-16.9. As it can be seen in Fig.~\ref{fig:lc}, the total brightness change of 
$0\fm22$ reported in filter B do not emerge over the scatter affecting AAVSO data.
However, the observations presented in Fig.~\ref{fig:lc} and in the top panel of 
Fig.~\ref{fig:lc_comp} show beyond any doubt that an outburst of AG~Peg occurred in the 
second half of 2015. During the first event the star brightness increased by 
$\sim\!2\fm2$ and $\sim\!1\fm8$ during the second, when compared to the mean values for 
B and V bands respectively. If compared to the first event,  the maximum of the second 
event was fainter by $\sim\!0\fm35$ in B filter and $\sim\!0\fm5$ in V filter.

\subsection{Spectroscopy}
\label{spec}

High-resolution ($\Delta R\sim\!30\,000$) spectra of AG~Peg, covering the spectral range 
from 4\,000 to 9\,000\,\AA, were obtained during 12 nights of the 
second half of 2015. We used the ESPERO echelle spectrograph (Bonev et al., in 
preparation) at the 2\,m RCC telescope of the Rozhen Observatory, Bulgaria. 
For all spectra S/N ratio is $\ge 50$ at 6\,000\,\AA. The spectra were 
reduced and calibrated using standard IRAF\footnote{\textsc{iraf} is distributed by the 
National Optical Astronomy Observatories, which are operated by the Association of 
Universities for Research in Astronomy, Inc., under cooperative agreement with the 
National Science Foundation.} procedures, including bias, dark, flat-field corrections 
and spectra extraction with background subtraction. 

Additionally, we used two spectra obtained with the ESO FEROS echelle 
spectrograph on October 2nd, 2004 and two spectra secured with the Asiago echelle 
spectrograph on August 13th and 14th, 2002. The FEROS spectral coverage extends from 
3\,600\,\AA\ to 9\,200\,\AA\ with resolving power of $\sim\!48\,000$. The Asiago spectra 
of AG~Peg are absolutely flux calibrated with $\Delta R \sim\!20\,000$ and cover the 
spectral region 3\,390-5\,985\,\AA. More details on FEROS and Asiago spectra can be found 
in \citet{2005MNRAS.363L..26Z} and \citet{2005A&A...434..397E} respectively. A journal of 
our spectral observations is presented in Table~\ref{tab:journal}.

\begin{table}
	\centering
	\caption{Journal of spectroscopic observation.}
	\label{tab:journal}
	\begin{tabular}{llll} 
		\hline
		Date & UT & Exp. time & Observat. \\
		 &  & [sec] & \\
		\hline
		2002 Aug 13 & 01:07 & 5670 & Asiago \\
		2002 Aug 14 & 00:53 & 1440 & Asiago \\
		2004 Oct 02 & 01:53 & 600  & ESO \\
		2004 Oct 02 & 01:42 & 600  & ESO \\
		2015 Jul 07 & 01:23 & 2820 & Rozhen \\
		2015 Jul 08 & 01:17 & 1320 & Rozhen \\
		2015 Aug 03 & 23:18 & 300  & Rozhen \\
		2015 Aug 04 & 22:19 & 300  & Rozhen \\
		2015 Oct 28 & 18:28 & 1800 & Rozhen \\
		2015 Oct 29 & 16:40 & 1800 & Rozhen \\
		2015 Nov 01 & 16:32 & 1800 & Rozhen \\
		2015 Dec 23 & 16:27 & 580  & Rozhen \\
		2015 Dec 24 & 16:09 & 160  & Rozhen \\
		2015 Dec 25 & 16:02 & 160  & Rozhen \\
		2015 Dec 26 & 16:22 & 160  & Rozhen \\
		2015 Dec 27 & 15:48 & 160  & Rozhen \\
		\hline
	\end{tabular}
\end{table}

We measured the equivalent widths (EW) of several emission lines integrating the area 
under the whole profile. Repeated test measurements have shown that the 
measurement error does not exceed 5\% for the strong and 10\% for the weak lines and it 
mainly depends on the local continuum level fitting. Then we calibrated the EWs of 
H$\beta$, \ion{He}{ii} 4686\,\AA\ and the Raman scattered line \ion{O}{vi} 6825\,\AA, 
measured in FEROS and ESPERO spectra, to fluxes. To do this, we used the AAVSO and 
\citet{2007AN....328..909S} BV photometry. To correct for the effect of emission 
lines, we reduced the flux in B by 30\% and in V by 10\%, adopting the values estimated 
by \citet{2007NewA...12..597S}. The fluxes were de-reddened with 
$E_\mathrm{B-V}$=0\fm1 adopted from \citet{1993AJ....106.1573K,2001AJ....122..349K} and 
using the standard ISM extinction curve of \citet{1999PASP..111...63F}. Considering the 
errors in the EWs and those added by the calibration, we estimate that the 
uncertainties in the fluxes can reach up to 20\% for the strong and up to 30\% for the 
weak emission lines.

The measured EWs and the fluxes of the emission lines are shown in 
Fig.~\ref{fig:lc_comp}. The emission line fluxes are also presented in 
Table~\ref{tab:th_lh}.

\section{Results and discussion}
\label{results}

\subsection{Changes in the spectrum during the outburst}
\label{sp_var}

\begin{figure}
	\includegraphics[width=\columnwidth]{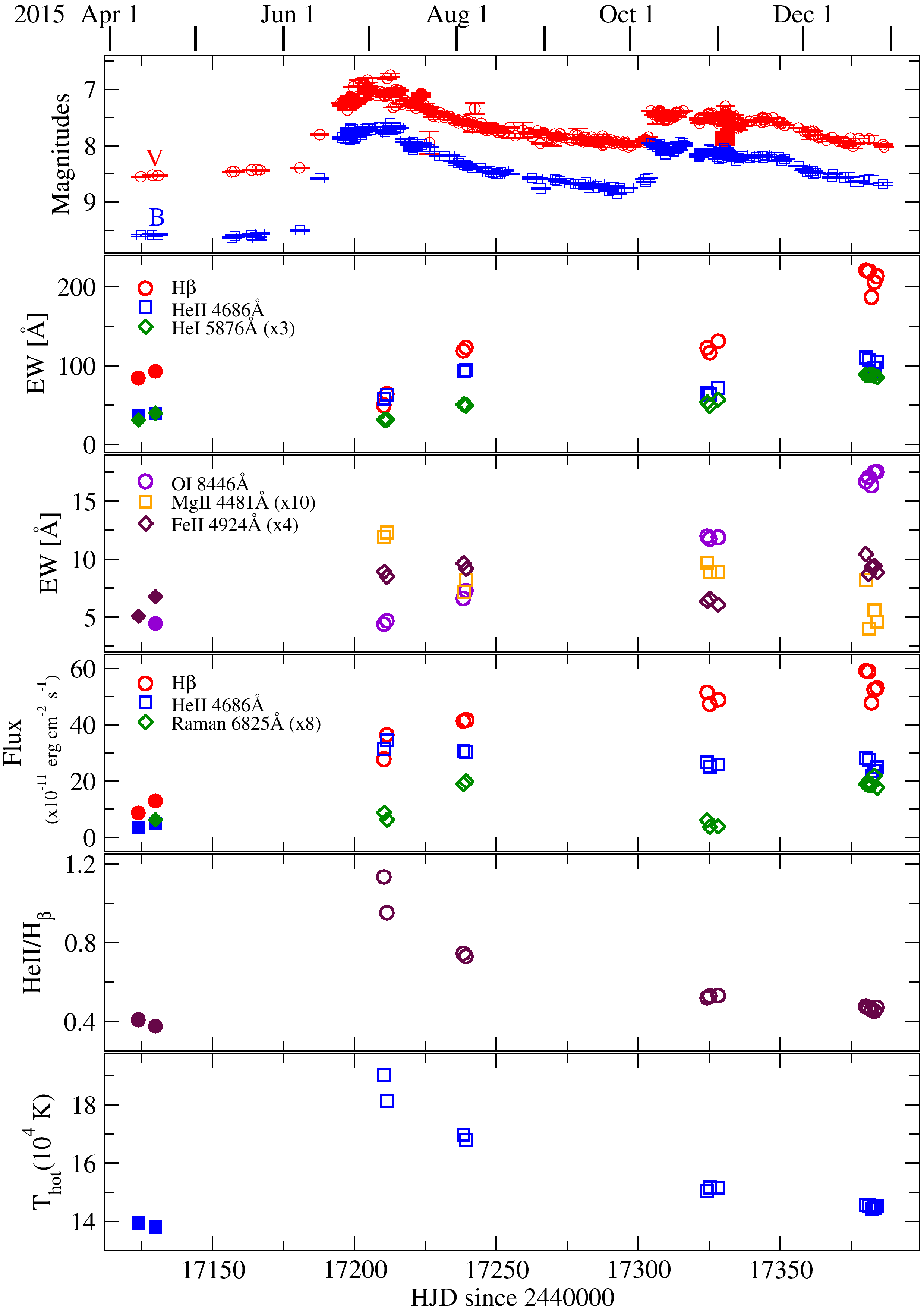}
    \caption{In the top panel an expanded view from Fig.~\ref{fig:lc} of the 
B (\textsl{squares}) and V 
(\textsl{circles}) light curves during the 2015 outburst of AG~Peg is shown. In the 
panels below the variations with the time of the EWs and fluxes for different spectral 
features are presented (see the text for details). In the two bottom panels the 
variations of the ratio \ion{He}{ii}\,4686\,\AA/H$\beta$ and the hot component 
temperature during the outburst are illustrated. The filled symbols in the lower left 
corner of each panel, excluding the top one, show only the respective values measured in 
the AG~Peg spectra obtained in 2002 (\textsl{left}) and 2004 (\textsl{right}). The 
figures in parentheses indicate how many times we have increased the respective value to 
show it in this scale.}
    \label{fig:lc_comp}
\end{figure}

The abrupt rise of AG~Peg optical brightness with $\sim$2 mag in 2015 is the first 
such event since its \revchanges{symbiotic nova outburst} in the mid-19th century. The 
first maximum occurred between July 2nd and 14th. The second took 
place around October 8th, followed by a plateau in the light curve  until around November 
24th. After that the star brightness decrease accelerated 
(Figs.~\ref{fig:lc} and \ref{fig:lc_comp}). The stages of the outburst, 
during which our spectra were obtained, are evident in Fig.~\ref{fig:lc_comp}. As 
Table~\ref{tab:th_lh} clearly shows, the October-December 2015 spectra and the spectra 
from Asiago and ESO, cover almost the same photometric phase.

\subsubsection{EWs and flux variations of different spectral lines}
\label{ew_flux}

\begin{figure*}
	\includegraphics[width=.75\textwidth]{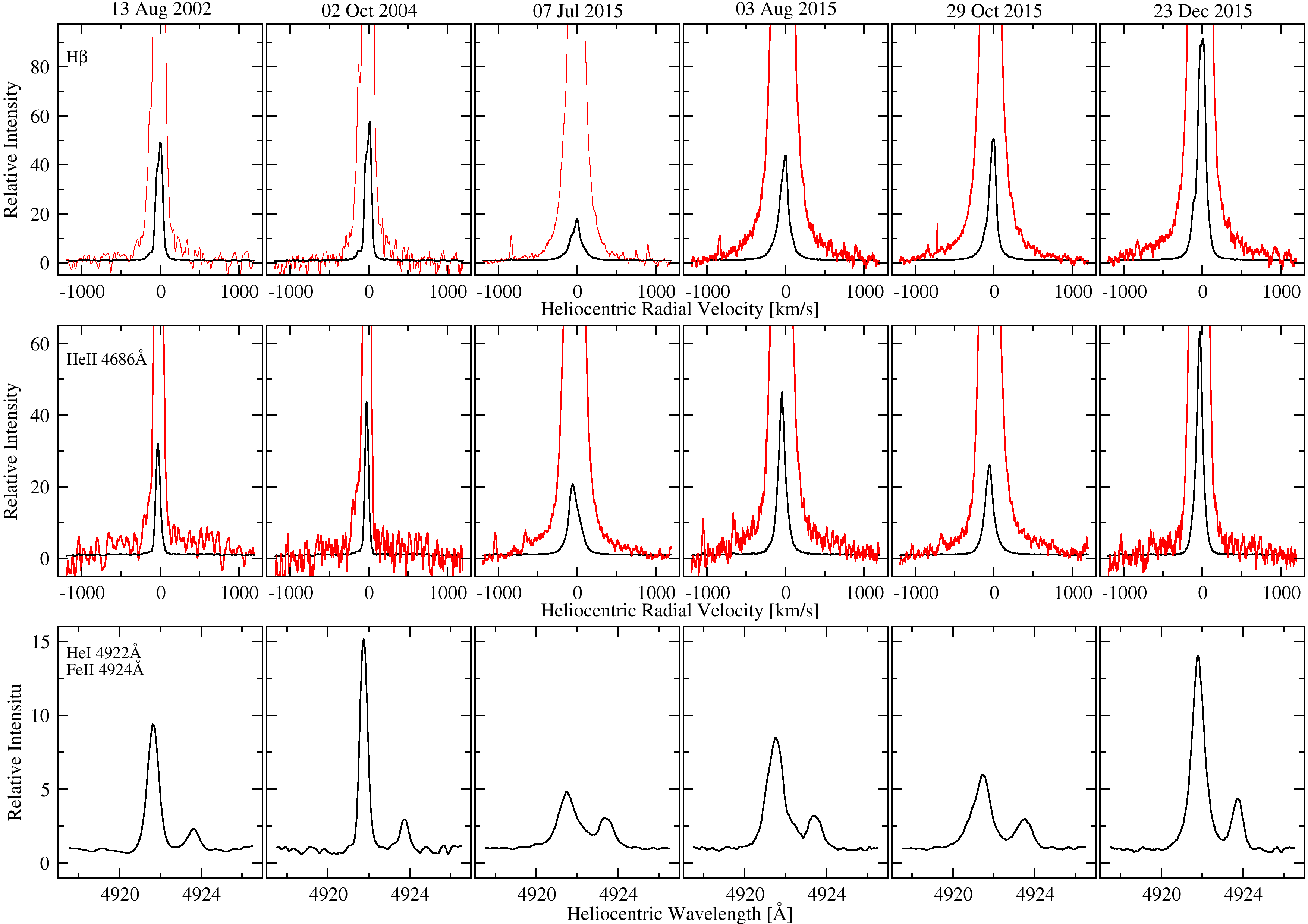}
    \caption{The remarkable changes in the profiles of different lines in the spectrum of 
AG~Peg are demonstrated. To make the broad emission wings well visible, with 
\textsl{thin} 
lines are represented the same profiles with relative intensity multiplied by 20.}
    \label{fig:profiles}
\end{figure*}

The increased brightness was accompanied by remarkable changes in the outburst 
spectrum of AG~Peg, when compared to the quiescence spectra in 2002 and 2004. A hot 
continuum considerably fills the cool star's absorption features, as it can be seen in 
the wide wings of \ion{He}{ii}\,4686\,\AA\ and H$\beta$ in Fig.~\ref{fig:profiles}. 
But, the A/F supergiant absorption spectrum, typical for symbiotic stars in 
outburst, is missing. Also, there were no signas of P~Cyg absorption components, 
indicating the existence of an expanding pseudo-photosphere, shell or massive, dense 
wind from the hot component. The most remarkable changes are in the intensity and the 
width of the emission lines as illustrated in Fig.~\ref{fig:profiles}. 

The FWHM of the \ion{He}{ii}\,4686\,\AA\ emission 
was 68\,\kms\ (2002) and 55\,\kms\ (2004) but in our spectra varied between 
107\,\kms\ and 169\,\kms. Similarly, the H$\beta$ FWHM from the range of 93\,\kms\ and 
97\,\kms\ increased, reaching the values of 120\,\kms\ and 183\,\kms.
The profiles of the Balmer lines, the \ion{He}{ii}\,4686\,\AA\ and 5412\,\AA\ and the most 
intensive \ion{He}{i} emissions are very complex and highly variable. In the outburst 
spectra of AG~Peg, shallow broad wings appeared in some of the strongest emission lines. 
The wings are typical in the case of classical symbiotics outburst. Their 
origin is still discussed, although different explanations, like for 
example, Raman scattering, electron scattering, optically thin wind from the hot 
component, etc. were suggested \citep[and references 
therein]{1989A&A...211L..27N,2000ApJ...541L..25L,2006A&A...457.1003S,2012BaltA..21..196S}.

The EWs of some strong emissions (Fig.~\ref{fig:lc_comp}) show growing trend, 
but with different gradients. H$\beta$ EW increased by a factor of ~4 while for 
\ion{He}{i}\,5876\,\AA\ and \ion{He}{ii}\,4686\,\AA\ the increase is of about 3 and 2 
times respectively. The \ion{Fe}{ii}\,4924\,\AA\ EW shows significant variations around 
the value $\sim\!2$\,\AA. The emission \ion{Mg}{ii}\,4481\,\AA, not presented in the 2002 
and 2004 spectra, was visible in all spectra obtained during the outburst, however, with 
an opposite trend of decreasing EW. The changes in the EWs, measured in our spectra from 
different epochs, are partly caused by the hot continuum variation. 
The \ion{O}{i}\,8446\,\AA\ EW grows almost linearly because the line is located 
relatively far to the red and is not affected by the hot continuum. The 
\ion{O}{i}\,7772\,\AA\ triplet is seen in absorption in the AG~Peg spectrum in 2004. It 
is visible in emission, 10 to 25 times fainter in comparison to the \ion{O}{i}\,8446\,\AA\ 
line, during the July-December 2015 period. Such large intensity of the emission 
\ion{O}{i}\,8446\,\AA\ with respect to the  \ion{O}{i}\,7772\,\AA\ triplet suggests 
fluorescence from Ly$\beta$ photons \citep{1947PASP...59..196B}.

We also show in Fig.~\ref{fig:lc_comp} the calculated fluxes of H$\beta$, 
\ion{He}{ii}\,4686\,\AA\ and the Raman scattered \ion{O}{vi} line 6825\,\AA.  
It is clear that the H$\beta$ flux almost doubled between July and December 
2015. During the same time, the \ion{He}{ii}\,4686\,\AA\ flux decreased by more than 
20\%.

As \citet{2001AJ....122..349K} demonstrated, the considerable
variations of the optical emission lines are correlated with the photometric phase. They 
show that the \ion{H}{i} and the \ion{He}{i} are more intensive roughly by a factor 2-3 
at photometric maxima in comparison to the minima. In the same time, the 
\ion{He}{ii}\,4686\,\AA\ intensity changes less than 50\%. The fluxes of H$\beta$ and 
\ion{He}{ii}\,4686\,\AA, around the maximum of photometric phase 18 (October-December 
2015), exceed more than 5-6 times the measured at the maximum of photometric phase 12, 
during the quiescence in 2002 and 2004. That is far above the accuracy ($\sim\!20\%$) of 
our measurements. This is evidence that the large increase of the intensities of the 
optical emission lines in 2015 spectra of AG~Peg is caused mainly by the outburst of this 
symbiotic system.

\begin{table*}
	\centering
	\caption{The de-reddened absolute fluxes of H$\beta$, \ion{He}{ii}\,4686\,\AA\ 
and 
the Raman scattered 6825\,\AA\ line, together with the estimated luminosity, temperature 
and radius of the AG~Peg hot component.}
	\label{tab:th_lh}
	\begin{tabular}{ccccccccc} 
		\hline		
\multicolumn{1}{c}{Date}&\multicolumn{1}{c}{HJD}&\multicolumn{1}{c}{Photom.}&\multicolumn{
3}{ c}{Flux $\times10^\mathrm{-11}$\,ergs\,cm$^\mathrm{-2}$\,s$^\mathrm{-1}$}& 
\multicolumn{1}{c}{$T_\mathrm{ h}$}&\multicolumn{1}{c}{$L_\mathrm{h}$} 
&\multicolumn{1}{c}{$R_\mathrm{h}$}\\ 
& \multicolumn{1}{c}{since 2440000}& 
\multicolumn{1}{c}{phase}&\multicolumn{1}{c}{\ion{He}{ii}\,4686\,\AA} & 
\multicolumn{1}{c}{H$\beta$} & \multicolumn{1}{c}{Raman 6825\,\AA} & 
\multicolumn{1}{c}{[10$^3$\,K]} & \multicolumn{1}{c}{[$L_{\sun}$]} & 
\multicolumn{1}{c}{[$R_{\sun}$]} \\
		\hline
2002 Aug 13 & 12499.55 & 11.99 & 3.64 & 8.90 &      & 140 & 240  & 0.027 \\
2002 Aug 14 & 12500.54 & 11.99 & 3.49 & 8.52 &      & 139 & 230  & 0.027 \\
2004 Oct 02 & 13280.07 & 12.95 & 4.89 & 12.9 & 0.85 & 138 & 340  & 0.033 \\
2004 Oct 02 & 13280.08 & 12.95 & 4.85 & 13.0 & 0.71 & 138 & 340  & 0.033 \\
2015 Jul 07 & 17210.53 & 17.76 & 31.5 & 27.8 & 1.09 & 190 & 1170 & 0.032 \\
2015 Jul 08 & 17211.55 & 17.76 & 34.5 & 36.3 & 0.78 & 181 & 1395 & 0.038 \\
2015 Aug 03 & 17238.47 & 17.79 & 30.7 & 41.2 & 2.38 & 170 & 1400 & 0.044 \\
2015 Aug 04 & 17239.43 & 17.79 & 30.4 & 41.7 & 2.49 & 168 & 1410 & 0.045 \\
2015 Oct 28 & 17324.26 & 17.90 & 26.7 & 51.4 & 0.76 & 150 & 1525 & 0.058 \\
2015 Oct 29 & 17325.19 & 17.90 & 25.1 & 47.4 & 0.47 & 152 & 1420 & 0.055 \\
2015 Nov 01 & 17328.18 & 17.90 & 25.9 & 48.8 & 0.49 & 152 & 1460 & 0.056 \\
2015 Dec 23 & 17380.18 & 17.97 & 28.2 & 59.1 & 2.37 & 146 & 1715 & 0.066 \\
2015 Dec 24 & 17381.17 & 17.97 & 27.6 & 58.8 & 2.33 & 145 & 1690 & 0.066 \\
2015 Dec 25 & 17382.16 & 17.97 & 21.9 & 47.7 & 2.35 & 144 & 1355 & 0.059 \\
2015 Dec 26 & 17383.18 & 17.97 & 23.7 & 52.5 & 2.74 & 145 & 1480 & 0.062 \\
2015 Dec 27 & 17384.15 & 17.97 & 24.9 & 53.0 & 2.21 & 145 & 1520 & 0.062 \\		
		\hline
	\end{tabular}
\end{table*}

\subsubsection{Raman scattered \ion{O}{vi} line 6825\,\AA}
\label{raman}

\begin{figure}
	\includegraphics[width=\columnwidth]{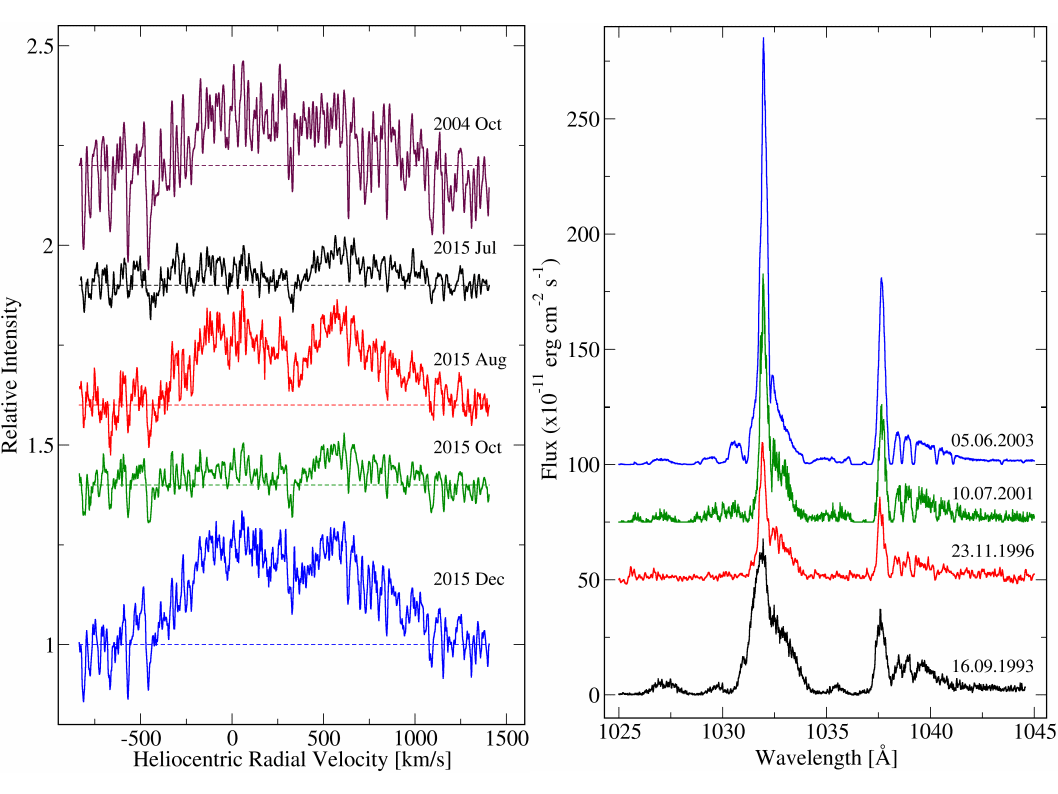}
    \caption{\textsl{Left}: The evolution of the Raman scattered \ion{O}{vi} 6825\,\AA\ 
emission line in our spectra of AG~Peg. The average, normalized to the local continuum 
profiles observed during the consecutive nights are shown. \textsl{Right}: The far UV 
emission lines \ion{O}{vi} 1032\,\AA\ and 1038\,\AA\ as observed by ORFEUS-I BEFS (1993), 
ORFEUS-II TUES (1996) and FUSE (2001 and 2003). The data was extracted from the STScI 
MAST archive in reduced format. The spectral fluxes are shifted for clarity by adding of 
different constant values.}
     \label{fig:ovi_all_6825}
\end{figure}

\citet{1999A&A...348..950S} subtracted the spectrum of a M3\,III giant to 
demonstrate the presence of the Raman scattered \ion{O}{vi} line 6825\,\AA\ 
\citep[see][]{1989A&A...211L..31S} in the spectrum of AG~Peg obtained in 1997. This line 
is directly visible in all our spectra obtained in 2004 and 2015 
(Fig.~\ref{fig:ovi_all_6825}). Partly because it is weaker and partly because of the 
molecular bands located around this wavelength, we were not able to identify the second 
Raman scattered \ion{O}{vi} line at 7082\,\AA\ in our spectra. The flux variations of the 
6825\,\AA\ emission are shown in Fig.~\ref{fig:lc_comp}. The changes are undoubtedly real 
because they exceed the $\sim\!30\%$ accuracy of the measurements.

In most symbiotic stars the Raman scattered \ion{O}{vi} lines exhibit a 
multi-peaked structure \citep[see][]{1999A&A...348..950S}. The \ion{O}{vi} 6825\,\AA\ 
emission profile in our spectra of AG~Peg has two components with similar peak 
intensities (Fig.~\ref{fig:ovi_all_6825}).

The minimal flux of the 6825\,\AA\ line measured by us 
(Table~\ref{tab:th_lh}) is more than three times greater than the reported by 
\citet{1999A&A...348..950S} 
$\sim\!1.4\pm0.7\times10^\mathrm{-12}$\,ergs\,cm$^\mathrm{-2}$\,s$^\mathrm{-1}$. This 
indicates that the 2015 UV emission \ion{O}{vi} 1032\,\AA\ should be stronger than that 
of the time of observations discussed by \citet{1999A&A...348..950S}. 
In Fig.~\ref{fig:ovi_all_6825}, we show all observations of the resonance doublet 
\ion{O}{vi} in the spectrum of AG~Peg existing in the MAST database. The data includes 
 ORFEUS (1993 and 1996) and FUSE (2001 and 2003) observations (for more details see 
\citet{1999A&A...348..950S} and \citet{2006A&A...451..157E}). A gradual increase of the 
intensity of the sharp component and a gradual decrease of the intensity of the broad 
wind component in the \ion{O}{vi} UV resonance lines are clearly apparent in 
Fig.~\ref{fig:ovi_all_6825}.

\begin{figure}
	\includegraphics[width=\columnwidth]{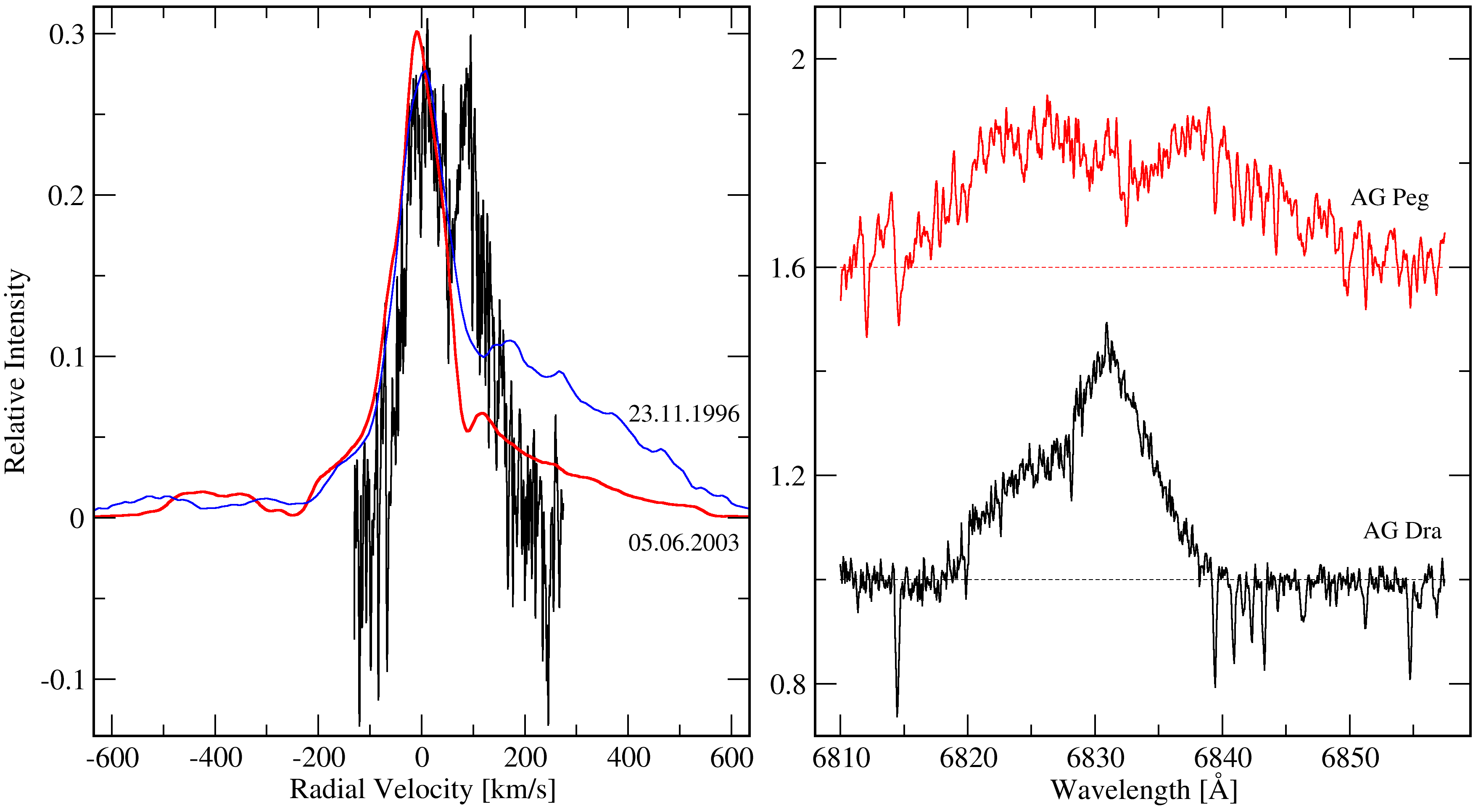}
    \caption{\textsl{Left}: Comparison of the Raman scattered \ion{O}{vi} 6825\,\AA\ line 
from our AG~Peg spectrum obtained in December 2015 with the original \ion{O}{vi} 
1032\,\AA\ emission as observed in 1996 (ORFEUS) and 2003 (FUSE). The Raman scattered 
line was converted to the original \ion{O}{vi} 1032\,\AA\ radial velocity scale. The 
fluxes of the \ion{O}{vi} 1032\,\AA\ emission were scaled in a way to be similar to the 
6825\,\AA\ line. All the profiles were smoothed with a moving box 7 pixels wide. 
\textsl{Right}: The profiles of the Raman scattered line 6825\,\AA\ in the spectra of 
AG~Peg and AG~Dra obtained with the ESPERO spectrograph in December and July 2015 
respectively.}
    \label{fig:raman_comp}
\end{figure}

In Fig.~\ref{fig:raman_comp}, we compare the mean Raman 
6825\,\AA\ profile (from our December 2015 spectra) and the \ion{O}{vi} line 1032\,\AA\ 
obtained by ORFEUS in 1996 and the most recent one obtained by FUSE in 2003). The 
6825\,\AA\ line profile is converted to the systematic radial velocity 
space of the original \ion{O}{vi} line 1032\,\AA\ emission using the coefficient 
$\lambda_\mathrm{Raman}/\lambda_{\ion{O}{vi}\,1032}=6.614$. The comparison 
shows that the \ion{O}{vi} line 1032\,\AA\ narrow profiles, as 
before \citep[see][]{1999A&A...348..950S}, well coincide with the blue component of the 
Raman scattered line at 6825\,\AA. Moreover, as visible in Fig.~\ref{fig:raman_comp}, 
there are no signs of the broad wind wings in the 6825\,\AA\ emission profile.

Also in Fig.~\ref{fig:raman_comp}, the mean Raman 6825\,\AA\ profile from our 
December 2015 spectra is compared with a typical 6825\,\AA\ scattered profile in the 
quiescence spectrum of AG~Dra \citep[][and references therein]{2010A&A...510A..70S} 
obtained with the same ESPERO spectrograph on July 5th, 2015. The most conspicuous are 
the remarkable broader profile in the AG~Peg spectrum and the difference in the blue and 
red components intensity ratio for both stars. 
Fitting two Gaussians to the AG~Peg Raman profile, we estimated the FWHM of the blue and 
red components to $400\pm40$\,\kms\ and $520\pm95$\,\kms\ respectively. For the total 
FWHM of the line we obtained the value of $1000\pm120$\,\kms which is slightly more than 
about $\sim\!800$\,\kms reported by \citet{1999A&A...348..950S} for the single component 
6825\,\AA. The calculated separation of the peaks is $614\pm62$\,\kms\ which converted 
to the systematic velocity scale of the \ion{O}{vi}\,1032\,\AA\ line gives 
$93\pm9$\,\kms. \citet{1999A&A...348..950S} and \citet{2010A&A...510A..70S} 
considered two different contributors to explain the double peaked Raman profiles in the 
spectrum of AG~Dra. The first one being the WD wind (and possibly accretion disk) and the 
second one - an ionized region surrounding it \revchanges{within the M star wind}. From 
the separation 
of the peaks of the line, \citet{2010A&A...510A..70S} estimated the scattering wind 
velocity to about 50\,\kms\ relative to the WD. If we treat the same way the 
Raman scattered 6825\,\AA\ line in the spectrum of AG~Peg than the wind velocity of 
$\sim\!100$\,\kms\ relative to the WD is twice as that for AG~Dra.

\citet{1999A&A...348..950S} interpreted the observed by them single 
peak AG~Peg Raman scattered 6825\,\AA\ profile as a strong support for the colliding wind 
model of \citet{1995A&A...297L..87M}. Results from several published observational 
studies (\citet[][and references 
therein]{1993AJ....106.1573K,2001AJ....122..349K,1994A&A...284..145V,2006A&A...451..157E}) 
provide evidence that the high velocity ($\sim\!1000$\kms), dense AG~Peg hot component 
wind was gradually decreasing during the last decades. The X-ray observations during the 
outburst in 2015 also do not confirm the presence of the dense and fast wind nor any 
noticeable signs of colliding stellar winds \citep{2016MNRAS.461..286Z}.

\citet{1999ApJ...515L..89L,2007ApJ...669.1156L,2015JKAS...48..105H} explain the two 
component profile of the Raman scattered lines observed in some symbiotics, proposing the 
existence of an accretion disk. In their model, the region emitting in UV \ion{O}{vi} 
resonance doublet is identified with the accretion flow around the WD and is divided 
into the blue and the red emission regions. In the framework of this model it is 
difficult to explain, why in the spectrum of AG~Peg first a single Raman scattered 
6825\,\AA\ emission was detected and only years later a second, red shifted component 
appeared.

\subsection{Hot component temperature and luminosity}
\label{th_lh}

To estimate the AG~Peg hot component temperature, the 
\citet{1981psbs.conf..517I} method was used, based on the emission line fluxes of 
\ion{He}{ii} 4686\,\AA, H$\beta$ and \ion{He}{i} 4471\,\AA, applicable to nebulae excited 
by a hot star and also optically thick to the hydrogen ionizing radiation. It is valid 
for effective temperatures of between 70\,000 and 200\,000\,K. A disadvantage of this 
method is that the lower members of the \ion{H}{i} Balmer series may be optically thick, 
which leads to an overestimate of the hot star effective temperature. Also, to some 
extent, it is sensitive to interstellar reddening 
\citep{1986syst.book.....K,1986A&A...155..137S}. To avoid the introduction of additional 
calibration errors, we used the equivalent widths instead of the line fluxes. 
\revchanges{The use of EWs for \ion{He}{ii} and H$\beta$
is based on their wavelength proximity, which nulls the effect of reddening
and slope of the underlying stellar continuum.} Because in 
our case the EW of \ion{He}{i} 4471\,\AA\ is always less than 10\% of both the 
\ion{He}{ii} 4686\,\AA\ and the H$\beta$ EWs, we neglected it using the 
EW$_\mathrm{4686}$/EW$_{H\beta}$ ratio. Since the EWs of emissions as strong as these of 
\ion{He}{ii} 4686\,\AA\ and H$\beta$ are measured with an error of the order of 5\%, the 
error of the calculated effective temperature should not exceed 15\%. The resultant 
effective temperature of the hot component is presented in Table~\ref{tab:th_lh} and its 
changes with the time are shown in Fig.~\ref{fig:lc_comp}.

We used de-reddened fluxes of \ion{He}{ii} 4686\,\AA\ and H$\beta$ to 
calculate the luminosity of the hot component in the same way as \citet[their 
equations 6 and 7]{1997A&A...327..191M}, adopting a distance of 800\,pc to AG~Peg 
\citep{1993AJ....106.1573K,2001AJ....122..349K}. To calculate the number of 
H$^\mathrm{0}$ and He$^\mathrm{+}$ ionizing photons, we used the \revchanges{number of 
ionizing photons $G_\mathrm{i}(T*)$} tabulated by \citet{1987A&A...182...51N}. In all 
cases, the 
differences between the luminosity calculated based on the \ion{He}{ii} 4686\,\AA\ and 
H$\beta$ do not exceed 20\%. For the hot component luminosity, presented in 
Table~\ref{tab:th_lh}, we adopted the average of the values obtained from both equations. 
We estimate the accuracy of the obtained luminosity to 20\%--30\%. But, considering the 
uncertainties in the reddening and the distance, the error in the luminosity could reach 
50\%.

The temperature and the luminosity of the AG~Peg hot component varied remarkably since 
its nova outburst. At the end of the 19th and the beginning of the 20th 
centuries, $T_\mathrm{h}$ was of the order of 8\,000--10\,000\,K and $L_\mathrm{h}$ of 
the order of 1\,000\,L$_{\sun}$. In the mid-20th century, the values changed to 
40\,000--50\,000\,K and 3\,000--3\,500\,L$_{\sun}$ respectively. Between 1970 and 1995, 
the temperature increased to $\sim\!95\,000-100\,000$\,K and luminosity dropped from 
$\sim\!1\,000-1\,200$\,L$_{\sun}$ to about 400\,L$_{\sun}$ \citep[see][and references 
therein]{1991A&A...248..458M,1993AJ....106.1573K,2001AJ....122..349K, 
1994A&A...282..586M,1997A&A...317..712A}.  

\citet{1995Obs...115..185Z} predict that the colliding winds stage 
will end, and accretion from the stellar wind will recommence, about the year 2001 and 
that the accretion will probably begin, when $L\simeq\!200$\,L$_{\sun}$. During the 
quiescence, we estimated a temperature of the hot component about 140\,000\,K and 
luminosity $\sim\!240$\,L$_{\sun}$ and $\sim\!340$\,L$_{\sun}$ for 2002 and 2004 
respectively (Table~\ref{tab:th_lh} and Fig.~\ref{fig:lc_comp}). Surprisingly, the 2002 
luminosity is practically identical to the value predicted by 
\citet{1995Obs...115..185Z}. 

Knowing the temperature and the luminosity of the hot component and supposing 
black body radiation, its radius can be estimated. The obtained values of 
$R_\mathrm{h}$ are listed in Table~\ref{tab:th_lh}. We estimated their accuracy to be of 
the order of 20\%--30\%. The radius of the AG~Peg hot component 
in 2002--2004 is about 0.03\,$R{\sun}$, which is in good agreement with the radii of the 
WDs of other symbiotic stars in quiescence \citep[see for 
example][]{1999PASP..111..571G}.

\subsection{The outburst}
\label{outburst}

To explain the classical symbiotic outbursts, different models were proposed: 
(i) WD photosphere expansion at constant bolometric luminosity, caused by an accretion 
rate exceeding that of steady burning \citep{1976Ap.....12..342T,1982ApJ...259..244I}; 
(ii) thermal pulse or shell flash \citep{1983ApJ...273..280K}; (iii) dwarf nova-like 
accretion disk instability 
\citep{1986A&A...163...56D,1986A&A...163...61D,2002ASPC..261..645M}; (iv) "combination 
nova" model  combining disk instabilities and enhanced thermonuclear shell burning 
\citep{2006ApJ...636.1002S}.

\begin{figure}
	\includegraphics[width=\columnwidth]{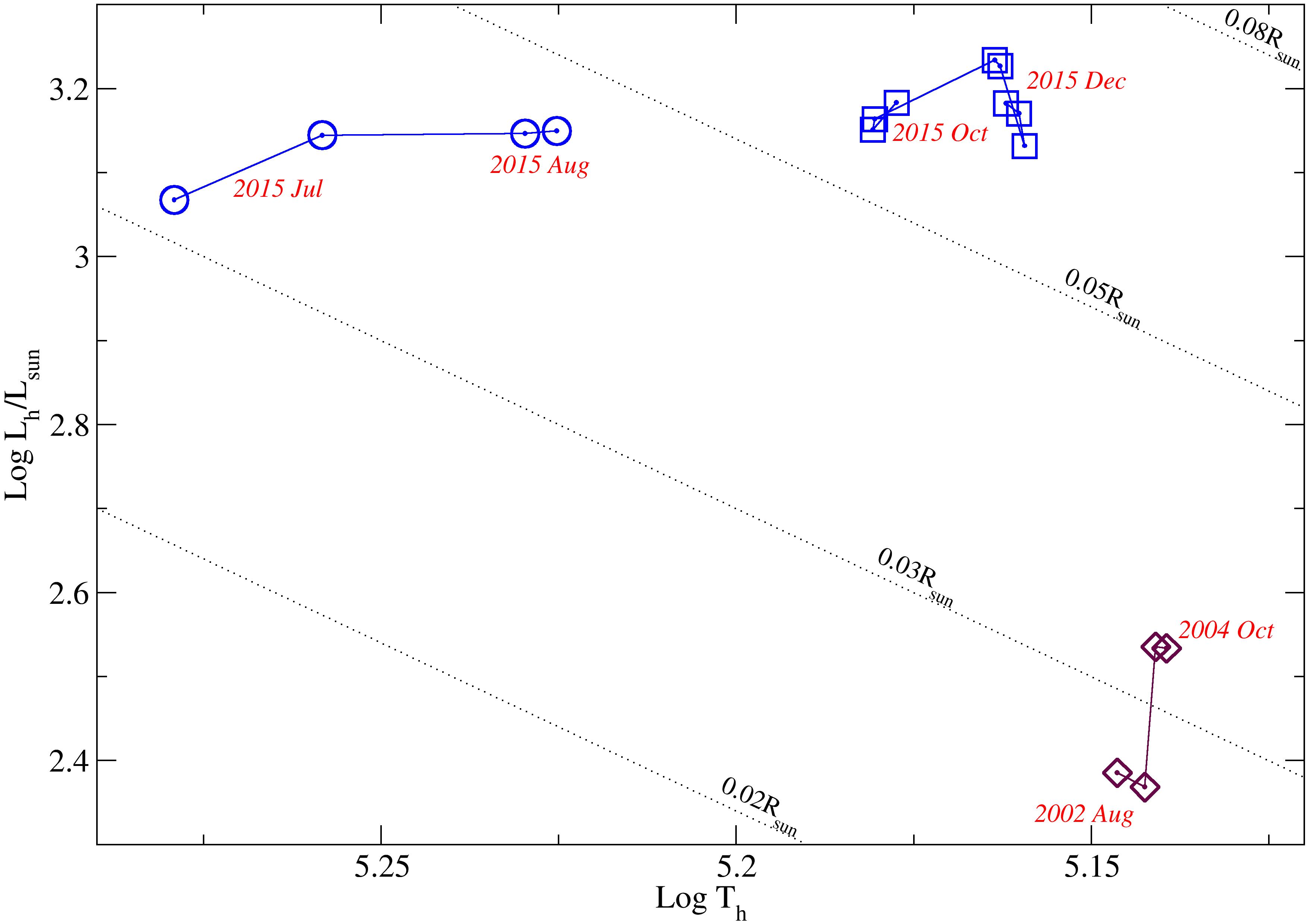}
    \caption{Evolution of the hot component of AG~Peg in the HR diagram.}
    \label{fig:wd_hr}
\end{figure}

As we mention before, a gradual decrease of the dense, high velocity wind of 
the AG Peg hot component was observed during the last decades. We also have not found 
signs of such a wind in our optical spectra obtained in 2002 and 2004.  Perhaps, 
\citet{1995Obs...115..185Z} were right and accretion from the red giant wind became 
possible around the year 2000. The hot component was accreting matter for about 15 years 
before the AG~Peg outburst took place in 2015. The recurrence time of a classical 
symbiotic outburst, with optical amplitudes of several magnitudes, is roughly a decade 
\citep{1986syst.book.....K}. The major outbursts of AG~Dra, for instance, occur on a 
time-scale 12--15\,yr \citep{2016MNRAS.456.2558L}.

Both AG~Peg brightness increases show steeper rise and slower decline. The 
second event started before the first one has ended. Similar in shape, overlapping 
maxima are observed during some of the high activity stages of AG~Dra and Z~And 
\citep{1999A&A...347..478G,2006ApJ...636.1002S,2008A&A...481..725G,2016MNRAS.456.2558L,
2009ApJ...690.1222S}.

The evolution of the AG~Peg hot component in the HR diagram during the 
outburst is shown in Fig.~\ref{fig:wd_hr}. The beginning of the outburst is marked by an 
increase of $T_\mathrm{h}$ while the hot component radius remained unchanged. In the 
month after the first brightness maximum, the hot component luminosity varied between 
1170\,L$_{\sun}$ and 1410\,L$_{\sun}$, without exceeding the estimated error. During the 
decline from the second maximum, between the end of October and the end of December, it 
was slightly higher in average, changing between 1355\,L$_{\sun}$ and 1715\,L$_{\sun}$ 
(again without exceeding the error). The temperature of the hot component decreased by 
25-30\% from July to December and the decline was steeper in the July-August period in 
comparison with the October-December period. The twofold increase of $R_\mathrm{h}$ in  
December in comparison with the first maximum in July is decisive for the observed high 
$L_\mathrm{h}$ during the decline from the second maximum. The strong increase of the hot 
component radius is, most probably, caused by a large expansion of the WD photosphere.
Similar changes in the hot component temperature and radius are typical for some of the 
outbursts of Z~And and AG~Dra \citep{2006ApJ...636.1002S,2008A&A...481..725G}.

The changes of the Raman scattered \ion{O}{vi} 6825\,\AA\ line in the 
spectrum of AG~Peg discussed in Section~\ref{raman} also resemble the behaviour of this 
line during the strong Z~And and AG~Dra outbursts. Numerous observations 
show that the Raman \ion{O}{vi} emission is weak or completely disappears around the 
maxima of these type of outbursts. 
\citep{2006ApJ...636.1002S,2009ApJ...690.1222S,2009PASP..121.1070M,2010A&A...510A..70S,
2016MNRAS.456.2558L}.

The results presented here, show that in some respects, the recent outburst of 
AG~Peg is similar to the strong outbursts of classical symbiotics like AG~Dra and Z~And. 
It is  hard to say what caused it: accretion on the WD surface, disk instability or 
thermonuclear shell burning on the WD surface triggered by disk instability.

\section{Conclusions}

Inspiration for this work was the behaviour of AG~Peg during the second half of 2015. 
We focus our studies on the outburst evolution using our own spectral 
observations, spectra obtained earlier in ESO and Asiago as well as archive data form 
AAVSO and MAST archives. Our main results and conclusions can be summarized as follows:

\begin{itemize}
\item[i)] The symbiotic nova outburst, affecting AG~Peg for more than a 
century, had ceased before the recent, different in nature, outburst took over in 2015.
\item[ii)] The earlier reported dense and massive high velocity wind from 
the hot component is not present anymore and accretion from the cool giant wind might 
have began around the year 2000. 
\item[iii)] Analysis of the evolution with time of the light curve, the EWs and the 
fluxes of different spectral features shows that in some respects the AG~Peg outburst 
resembles those observed in the classical symbiotic stars like Z~And and 
AG~Dra. 
\item[iv)] A strong argument in favour of the similarity of the AG~Peg 
outburst to the classical symbiotic outbursts is the observed behaviour of the 
Raman scattered \ion{O}{vi} emission line at 6825\,\AA.
\end{itemize}
In conclusion, it could be said that the recent photometric and spectral variations of 
AG~Peg place it among the classical Z~And type symbiotic binary systems.

\section*{Acknowledgements}

We acknowledge with thanks the variable star observations from the AAVSO International 
Database contributed by observers worldwide and used in this research.
This research has made use of the NASA’s Astrophysics Data System, and the SIMBAD 
astronomical data base, operated by CDS at Strasbourg, France. Some of the data 
presented in this paper were obtained from the Mikulski Archive for Space Telescopes 
(MAST). STScI is operated by the Association of Universities for Research in Astronomy, 
Inc., under NASA contract NAS5-26555. Support for MAST for non-HST data is provided by the 
NASA Office of Space Science via grant NNX09AF08G and by other grants and contracts. We 
are thankful to U. Munari who made available the Asiago spectra of AG~Peg obtained in 
2002. We are grateful to the anonymous referee for valuable comments and 
suggestions. TT is very grateful to Svet Zhekov for the many useful discussions 
and Sz. Zywica for some help with English.




\bibliographystyle{mnras}
\bibliography{t_tomov} 






\bsp	
\label{lastpage}
\end{document}